\documentclass[11pt]{article}
\usepackage{hyperref}
\begin{document}
\title{Visualization of Word Usage in Fluid Mechanics Abstracts}
\author{Eric Mockensturm and Kendra Sharp \\ Penn State University \\University Park, PA, 16802, USA}
\maketitle
\vspace{-8mm}

The fluid dynamics video shows how `keywords' from abstracts contained in three journals--Physics of Fluids (A) from 1970, Experiments in Fluids from 1983, and Journal of Fluid Mechanics from 1954--have changed over time.  Abstracts were analyzed by first removing non-descriptive words.  A script then parsed each abstract to find all unique words that remained.  Each abstract was then analyzed to determine if a given word appeared in the abstract.  For each year the twenty words that appeared in the most abstract were recorded.  The union of these top twenty words ($\sim$50 words) was then used to study the abstracts over all the years.  The percentage of abstracts in which each word occurred was then calculated for each year studied.  In this way, usage trends of common descriptive words over the course of many years are extracted.  The entire process, from collecting abstracts to  extract the frequency of word use, is automated so that any collection of abstracts, titles, or keywords can be easily analyzed.

The data is visualized and animated using an applications written by the authors.  Each word is represented by a bubble whose size represents the percentage of abstracts containing the word.  This program uses Apple Computer's Quartz and CoreAnimation technologies for rendering and a high-order numerical integrator to determine particle motion.  Word bubbles are modeled as soft spheres, repelling each other when they overlap and having a mass proportional to the square of their radius.  They are attracted to the center of the domain by nonlinear springs and have a vertical downward (gravitational) force such the larger bubbles tend to sink to the bottom of the region.  The center of mass of each bubble is offset from the geometric center of the circle through which the spring and contact forces act.  Each bubble then behave like a pendulum with sprung base.  Damping is included so the motion settles into quasi-period behavior.

The associated video can be viewed at \href{http://hdl.handle.net/1813/14072}{eCommons@Cornell}.
\end{document}